\documentclass[twocolumn,a4paper,10pt,reprint,aip]{revtex4-1}

\usepackage{graphicx}
\usepackage{latexsym}   % useful for coding complex math
\usepackage{amsmath,amssymb,amsthm}

\usepackage{upgreek}
\usepackage{pifont}
\usepackage{mathrsfs}
\usepackage{isomath}
\usepackage{xcolor}

\usepackage[hidelinks]{hyperref}
\hypersetup{
    colorlinks = true,
    citecolor = blue,
    linkcolor=blue
}

\makeatletter
\renewcommand{\p@subsection}{}
\renewcommand{\p@subsubsection}{}
\makeatother

\begin{document}

\title{Microrheological model for Kelvin-Voigt materials with micro-heterogeneities}

\author{T. N. Azevedo}
\affiliation{Departamento~de~F\'isica,~Universidade~Federal~de~Vi\c{c}osa~(UFV),~36.570-900,~Vi\c{c}osa,~MG,~Brazil.}

\author{K. M. Oliveira}
\affiliation{Departamento~de~F\'isica,~Universidade~Federal~de~Vi\c{c}osa~(UFV),~36.570-900,~Vi\c{c}osa,~MG,~Brazil.}
 
\author{H. P. Maia}
\affiliation{Departamento~de~F\'isica,~Universidade~Federal~de~Vi\c{c}osa~(UFV),~36.570-900,~Vi\c{c}osa,~MG,~Brazil.}

\author{A. V. N. C. Teixeira}
\affiliation{Departamento~de~F\'isica,~Universidade~Federal~de~Vi\c{c}osa~(UFV),~36.570-900,~Vi\c{c}osa,~MG,~Brazil.}

\author{L. G. Rizzi}
\affiliation{Departamento~de~F\'isica,~Universidade~Federal~de~Vi\c{c}osa~(UFV),~36.570-900,~Vi\c{c}osa,~MG,~Brazil.}

%%%%% ABSTRACT %%%%%
\begin{abstract}
  We introduce a generalization of the Kelvin-Voigt model in 
order to include and characterize heterogeneities in viscoelastic semisolid 
materials. 
  By considering a microrheological approach, we present analytical expressions
for the mean square displacement and for the time-dependent diffusion coefficient 
of probe particles immersed in a viscoelastic material described by this model.
  Besides validating our theoretical approach through Brownian dynamics simulations,
we show how the model can be used to describe experimental data obtained for
polyacrylamide and laponite gels.
\end{abstract}

\maketitle

\section{Introduction}
\label{intro}

  Although many viscoelastic materials are observed to have heterogeneous 
structures~\citep{graessley_2003}, {\it e.g.}, agarose~\citep{Valentine2001}, 
actin~\citep{Apgar2000}, collagen~\citep{Shayegan2013}, hectorite dispersions~\citep{donald2008eurphysJE}, 
fibril networks~\citep{donald2009eurphysJE} and peptide gels~\citep{donald2012softmatter,donald2014eurphysJE}, 
their theoretical characterization has been the subject of a rather limited number of 
studies~\citep{krall1998prl,Savin2007,rizzi2020jrheol}.
  Extensive experimental research has shown that heterogeneities,
in addition to influencing the formation kinetics of complex materials~\citep{Tseng2002,Tseng2004},
can play an important role in their rheological and optical properties~\citep{Quintanilla1999}.

	Among the various mechanical models of viscoelasticity, the Kelvin-Voigt (KV)
model is one of the simplest theoretical models that can be used to describe viscoelastic 
materials which display a semisolid response~\citep{ferrybook,raobook}.
	Basically, it can be characterized by a complex modulus $G^{*}(\omega)$ given by
\begin{equation}
  G^{*}(\omega)=G_0+i\eta_0\omega ~,
	\label{kv_model_modulus}
\end{equation}
where $G_0$ is the storage modulus and $\eta_0$ is the viscosity of the material.
	Although idealized, the KV model has been used to describe the viscoelasticity
of several semisolids, and examples include	fibrin networks~\citep{garcia2020}, 
gelatin hydrogels~\citep{SHABANIVERKI2017}, polyurea~\citep{NANTASETPHONG2016142}, 
and laponite gels~\citep{rich2011}.
	As it happens, a detailed look at the results generally indicates that Eq.~\ref{kv_model_modulus}
works just as a rough approximation and it might be that the departures from such an ideal behaviour
could involve, for instance, the presence of heterogeneities in the sample~\citep{donald2008eurphysJE}.

	Here we consider a microrheology-based approach to propose a generalization of the KV model 
in order to incorporate the micro-heterogeneities present in semisolid viscoelastic materials.
	In particular, by assuming that the micro-heterogeneities can be described by position-dependent 
spring constants $\varepsilon=\varepsilon(\vec{R})$ and drag coefficients $\nu = \nu(\vec{R})$,
we are able to describe analytically the power-law behaviour of the time-dependent diffusion 
coefficient $D(\tau)$ at long times, and relate its exponent to the distribution of spring constants 
and drag coefficients.
	Accordingly, we perform stochastic simulations using Brownian dynamics to validate the expressions
obtained for $D(\tau)$ and for the mean-squared displacement.
	Then, we show how our model can be used to describe experimental data obtained from dynamic light
scattering (DLS) microrheology for two different kinds of gels.

	The manuscript is organized as follows. 
	First, in Sec.~\ref{KV_section}, we review some aspects of the microrheology of viscoelastic materials
that exhibit a Kelvin-Voigt type behaviour (Eq.~\ref{kv_model_modulus}).
	In Sec.~\ref{KVMH_section}, we present the theoretical description of the model introduced here, 
the Kelvin-Voigt with micro-heterogeneities (KVMH) model, which generalizes the usual KV model.
	Section~\ref{numerical_sim} presents the validation of our theoretical results through numerical 
simulations using Brownian dynamics.
	In Sec.~\ref{Experimental_section} we apply our model in the analysis of data obtained from 
microrheology experiments, and we show a comparison between the two models to emphasize how the
KVMH might be more suitable to describe experimental results when micro-heterogeneities are present in 
the sample.

%========================================================================
%========================================================================
\section{Microrheology of semisolid Kelvin-Voigt (KV) materials}
\label{KV_section}
%========================================================================
%========================================================================

	Before taking into account the micro-heterogeneities, we briefly review how one can consider an 
analytical approach to describe the microrheological response of the usual Kelvin-Voigt material.
	In particular, by considering the linear viscoelastic (LVE) regime~\citep{Rizzi_Tassieri_2018},
one has that the mechanical response of a viscoelastic medium can be probed by the stochastic movement 
of spherical particles moving through the sample,	which are characterized by their mean-squared 
displacement (MSD) $\langle \Delta r^2(\tau) \rangle \equiv \langle [\vec{r}(\tau) -  \vec{r}(0) ]^2 \rangle$.
	Experimentally, the MSD of probes can be obtained through, {\it e.g.}, particle tracking videomicroscopy
or dynamic light scattering techniques~\citep{Waigh_2016}, so that, by considering the following generalized
Stokes-Einstein relationship (GSER), one can obtain the compliance of the material 
as~\citep{Squires_Mason_2016_Review,Mason2000}
\begin{equation}
  J(\tau) = \frac{3\pi a}{ d_e k_BT}\langle \Delta r^2(\tau) \rangle ~~,
  \label{compliance}
\end{equation}
where $k_B$ and $T$ are the Boltzmann's constant and the absolute temperature of the medium, respectively;
$a$ is the radius of the probe particles, and $d_e$ is the Euclidean dimension of the random walk.
	Importantly, in this work we will only consider disordered and isotropic materials so that the compliance
$J(\tau)$ is a scalar function of time.
	In addition, it is worth noting that the above GSER is derived under the assumption that the radii of the
probe particles are larger than the largest microstructures of the material,and also that the inertia of the 
probe particles can be neglected~\citep{Waigh_2016}.
	Under these conditions, the results obtained by the microrheology technique should reproduce the results 
obtained from bulk rheology.
	Hence, through the Fourier-Laplace transform of $J(\tau)$, {\it i.e.}, 
$\hat{J}(\omega) = \mathcal{L}[J(\tau);s]_{s=i\omega}$, one can obtain the complex modulus 
$G^{*}(\omega) = G^{\prime}(\omega) + i G^{\prime \prime}(\omega)$ of the material through the following 
identity~\citep{Waigh_2016,Rizzi_Tassieri_2018}
\begin{equation}
  G^{*}(\omega) = \frac{1}{i\omega \hat{J}(\omega)} ~~.
  \label{complemodulus} 
\end{equation}
	Numerically, one can compute the Fourier-Laplace transform, $\hat{J}(\omega)$, by using the method proposed
in Ref.~\citep{Evans2009}.

	Next, for simplicity, we consider a one-dimensional ($d_e=1$) stochastic movement of a Brownian particle 
trapped in a harmonic potencial, $U(x)=\kappa x^2/2$, so that it can be described by the following overdamped 
Langevin equation,
\begin{equation}
  \zeta \frac{ d }{d t} x(t) = - \frac{d}{dx} U(x)  + f_{a}(t) ~~,
\end{equation}
where $\zeta$ is the drag coefficient, $f_{a}(t)$ is a random force that the surrounding medium exerts on the
particle; here the constant $\kappa$ denotes the strength of a ``spring constant'' which mimics the properties
of the viscoelastic material and characterizes its solid-like response.
	As it is shown in Ref.~\citep{azevedo2020jphysconfser}, the solution of the above overdamped Langevin equation 
yields a mean-squared displacement which corresponds to that of a probe particle immersed in a Kelvin-Voigt 
material, that is,
\begin{equation}
  \langle \Delta x^2(\tau) \rangle = \frac{C}{2 \gamma } \left[ 1 - \exp{ (-2 \gamma \tau) }  \right]~,
  \label{deltar2}
\end{equation}
where $\gamma=\kappa/\zeta$.
	As discussed in Ref.~\citep{gillespie1993amjphys}, the fluctuation-dissipation theorem links the constant $C$ 
to the auto-correlation function of the ``forces'', $\tilde{f}_a = f_a/\zeta$, which is given by 
$\langle \tilde{f}_a(\tau) \tilde{f}_a(\tau^{\prime})\rangle = C \delta(\tau - \tau^{\prime})$.
	By considering that $\langle \Delta x^2(\tau) \rangle = \langle x^2(\tau) \rangle$ with $x(0)=0$, and assuming
that the equipartition theorem holds, one has that 
$\kappa \langle \Delta x^2(\infty) \rangle = \kappa \langle x^2(\infty) \rangle = k_BT$, which leads to 
$C=2 \, k_BT / \zeta$.
	That value is consistent with a force auto-correlation function that is given by
$\langle f_a(\tau) f_a(\tau^{\prime})\rangle = 2 k_BT \zeta \, \delta(\tau - \tau^{\prime})$.

	Hence, from Eq.~\ref{deltar2}, one finds that the MSD of the usual KV is given by
\begin{equation}
  \langle \Delta x^2(\tau) \rangle =
    \frac{k_BT}{\kappa} \left[ 1 - \exp{ \left(- 2 \frac{\tau}{\tau_c} \right) } \right]~,
  \label{deltar2_KV_uni} 
\end{equation}
where $\tau_c=(\gamma)^{-1}$ is a characteristic ({\it i.e.}, corner) time. 
	Accordingly, Eq.~\ref{deltar2_KV_uni} displays a linear ({\it i.e.}, normal diffusive) behaviour
$\langle \Delta x^2(\tau) \rangle \approx 2 (k_BT / \zeta) \tau$ for $\tau \ll \tau_c$, and a constant value,  
$\langle \Delta x^2(\tau) \rangle \approx k_BT / \kappa$, when $\tau \gg \tau_c$, indicating that 
the displacements of the probe particle are restricted to a region of the sample at long times.
	In addition, one can evaluate the time-dependent diffusion coefficient as 
\begin{equation}
  D(\tau)   =  \frac{1}{2 d_e}\frac{d}{d\tau} \langle \Delta r^2(\tau) \rangle = 
    \frac{k_B T}{\kappa\tau_c}\exp\left( -2\frac{\tau}{\tau_c} \right)~,
  \label{D_KV_final}
\end{equation}
where the right side of the above identity does not depend on $d_e$ since 
$\langle \Delta r^2(\tau) \rangle=d_e\langle \Delta x^2(\tau) \rangle$ for isotropic materials.
	Now, by considering the GSER given by Eq.~\ref{compliance}, one can use Eq.~\ref{deltar2_KV_uni} to obtain 
the compliance $J(\tau)$, which can be readily identified as the compliance of a KV material~\citep{ferrybook}.
	As shown in Ref.~\citep{azevedo2020jphysconfser}, one can use the corresponding compliance $J(\tau)$ 
to evaluate both the storage, $G^{\prime}(\omega)$, and the loss modulus, $G^{\prime \prime}(\omega)$, of the 
material through Eq.~\ref{complemodulus}, that is,
\begin{equation}
  G^{*}(\omega) = G^{\prime}(\omega) + i G^{\prime \prime}(\omega) = 
    \frac{\kappa}{3 \pi a} + i \frac{\zeta}{6 \pi a}  \omega ~~.
\end{equation}
  By comparing this last expression with Eq.~\ref{kv_model_modulus}, one can identify $G_0=\kappa/3\pi a$ and
$\eta_0=\zeta/6\pi a$. 
	We note that, since the measured moduli should not exhibit any dependence on the radius $a$ of the probe 
particle, both $\kappa=\kappa(a)$ and $\zeta=\zeta(a)$ are expected to display a linear dependence on $a$.
  Thus, based on the microrheology of a KV material, one should have that the spring constant
and the drag coefficient are respectively related to the plateau modulus and the viscosity of the KV material, 
that is, $\kappa = 3\pi a G_0$ and $\zeta = 6\pi a \eta_0$.

%-%-%-%-%-%-%-%-%-%%-%-%-%-%-%-%-%-%-%
%-%-%-%-%-%-%-%-%-%%-%-%-%-%-%-%-%-%-%
%-%-%-%-%-%-%-%-%-%%-%-%-%-%-%-%-%-%-%
\begin{figure}[!t]
\centering
\includegraphics[width=0.48\textwidth]{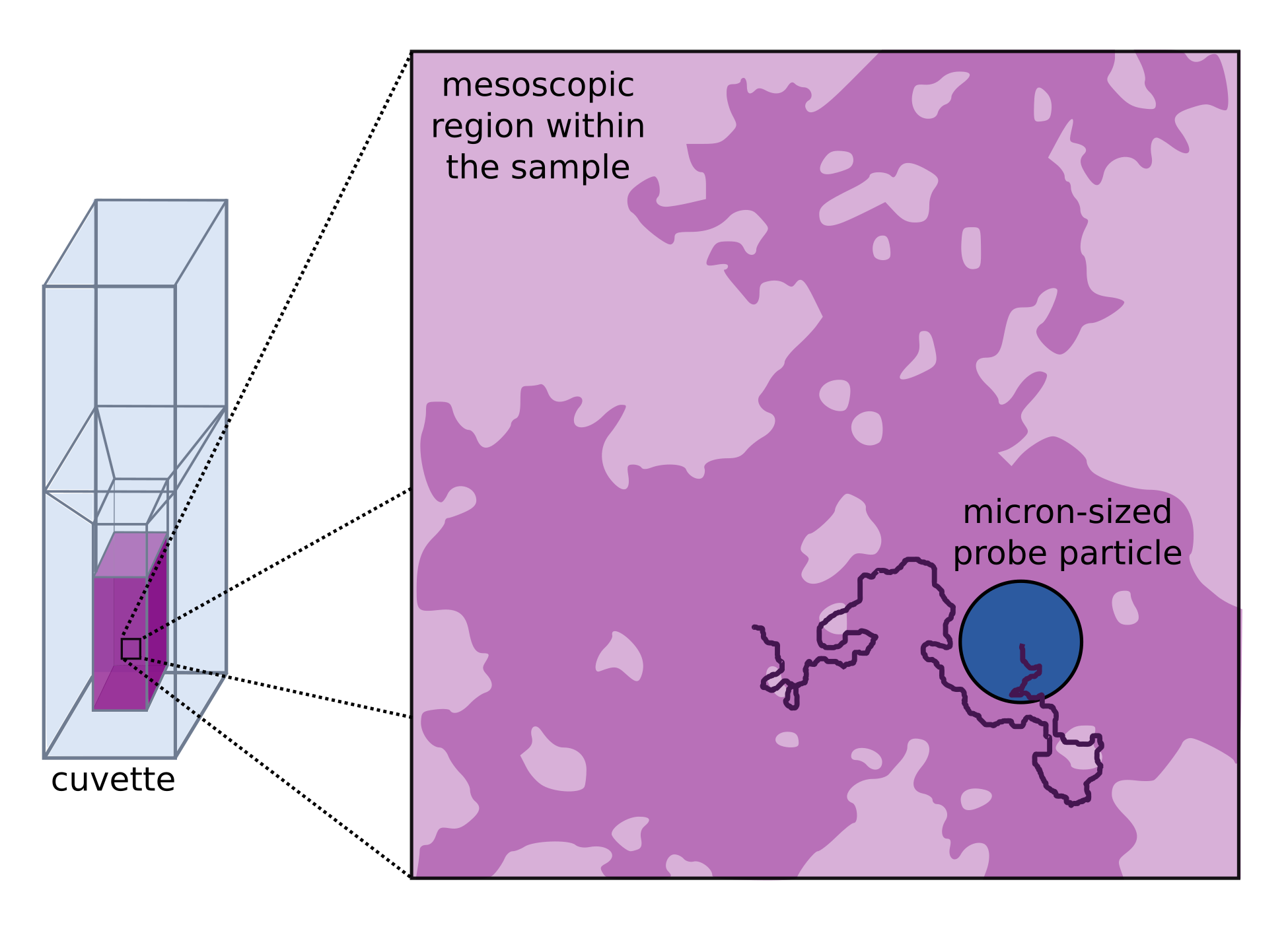}
\caption{Illustration of the heterogeneous microrheology that is described by the KVMH model, where a
single micron-sized probe particle immersed in a semisolid experiences a viscoelastic response that is
characterized by a spring constant $\varepsilon(\vec{R})$ and a drag coefficient $\nu(\vec{R})$, which
are locally-defined near a mesoscopic region denoted by $\vec{R}$.}
\label{referenceframe}
\end{figure}
%-%-%-%-%-%-%-%-%-%%-%-%-%-%-%-%-%-%-%
%-%-%-%-%-%-%-%-%-%%-%-%-%-%-%-%-%-%-%
%-%-%-%-%-%-%-%-%-%%-%-%-%-%-%-%-%-%-%

%========================================================================
%========================================================================
\section{Kelvin-Voigt model with micro-heterogeneities (KVMH)}
\label{KVMH_section}
%========================================================================
%========================================================================

	Now, in order to take into account the micro-heterogeneities, we assume that the sample has a region-dependent
drag coefficients $\nu = \nu(\vec{R})$,	where $\vec{R}$ is the position of a mesoscopic region within the sample,
as illustrated in Fig.~\ref{referenceframe}.
	Although the elastic properties in different regions of the system are heterogeneous throughout the sample, we 
assume that, locally, the drag coefficient is constant within a given mesoscopic region  so that the probe 
particle will not experience drastic changes in the viscoelastic properties of the medium while being in that 
region.
	Also, because the probe particle is trapped by a harmonic potential, which is defined by the local spring 
constant $\varepsilon=\varepsilon(\vec{R})$, it is not expected to leave that mesoscopic region during the 
typical time scale of the experiments $\tau_e \gg \tau_c$.

	In the following, we discuss both the theoretical and computational aspects involved in the description of the KVMH.

%========================================================================
\subsection{MSD and the distribution of micro-heterogeneities}
\label{Theory_section}
%========================================================================

	Following the approach introduced in Ref.~\citep{rizzi2020jrheol}, we assume that the MSD (i.e., averaged over 
trajectories of probe particles which experience different viscoelastic properties due to the micro-heterogeneities)
can be evaluated as
\begin{equation}
  \langle \Delta x^2(t) \rangle = \int  \langle \Delta x^2(t) \rangle_\xi \, \rho(\xi) \,d\xi ~~,
  \label{final_MSD}
\end{equation}
where the locally defined microrheological properties of mesoscopic regions within the system  are incorporated into
the model through the distribution $\rho(\xi)$.
	As in Ref.~\citep{rizzi2020jrheol}, here we define $\rho(\xi)$ as a gamma distribution~\citep{Crooks2019}, that is,
\begin{equation}
  \rho(\xi) = \frac{\xi^{-(1-p)}e^{-\xi}}{\Upgamma(p)}~~,
  \label{rho_dist}
\end{equation}
with $\Upgamma(p)$ being the usual gamma function and $p$ is an exponent that characterizes the distribution of the
region-dependent variable $\xi=\xi(\vec{R})$, so that the mean value of $\xi$ is given by
\begin{equation}
  \bar{\xi}= \int_{0}^{\infty} \, \xi \, \rho(\xi) \, d\xi = p ~~.
  \label{mean_xi}
\end{equation}
  In particular, by considering that $\gamma_{\xi}^{\,} = \kappa/\nu_{\xi}$, with the local drag coefficient given by
\begin{equation}
  \nu_{\xi}^{\,} \equiv \frac{p}{\xi}\,\zeta~~,
  \label{nu-xi=p_xi_zeta}
\end{equation}
it follows from the analogy with Eq.~\ref{deltar2} that
\begin{equation}
  \langle \Delta x^2(\tau) \rangle_{\xi} = \frac{C_{\xi} \nu_{\xi}}{2\kappa} 
  \left[ 1 - \exp{ \left(- 2 \frac{\kappa}{\nu_{\xi}^{\,}} \tau \right) }  \right]~~,
  \label{deltar2_xi}
\end{equation}
where the constant $C_{\xi}$ can be estimated by assuming that the equipartition theorem is locally valid, that is,
\begin{equation}
  C_{\xi}=2 \left( \frac{k_BT }{\zeta} \right) \left( \frac{\xi}{p}  \right)~~,
\end{equation}
with $\xi$ distributed according to Eq.~\ref{rho_dist}.
	Interestingly, for short times, Eq.~\ref{deltar2_xi} yields 
$\langle \Delta x^2(\tau) \rangle_{\xi} \approx  C_\xi \tau$, so that Eq.~\ref{final_MSD} together with 
Eq.~\ref{mean_xi} leads to the usual linear diffusive behaviour, that is,
\begin{equation}
  \langle \Delta x^2(\tau) \rangle \approx 
    2 \left( \frac{k_B T}{\zeta} \right) \tau~~.
\label{short-times-final}
\end{equation}
	On the other hand, at later times, Eq.~\ref{deltar2_xi} yields
\begin{equation}
  \langle \Delta x^2(\tau) \rangle_{\xi} \approx \frac{C_{\xi} \nu_{\xi}}{2 \kappa} =
    \frac{k_BT}{\kappa}~~,
\label{later-times}
\end{equation}
so that, because the distribution $\rho(\xi)$ is normalized, Eq.~\ref{final_MSD} also leads to a $\xi$-independent
result for the total MSD, that is,
\begin{equation}
  \langle \Delta x^2(\tau) \rangle \approx \int_{0}^{\infty}   \left( \frac{k_B T}{\kappa} \right) \, \rho(\xi)\,d\xi
    = \frac{k_B T}{\kappa} ~~.
  \label{later-times-final}
\end{equation}
	Obviously, by considering $\rho(\xi)$ given by Eq.~\ref{rho_dist}, one can evalute the expression for the MSD
more generally from Eq.~\ref{deltar2_xi}.
	Indeed, by performing the corresponding integral~\citep{gradshteyn} one finds from Eq.~\ref{final_MSD} that
\begin{equation}
  \langle \Delta x^2(\tau) \rangle = \frac{k_B T}{\kappa} \left[1-\left(\frac{2\tau}{p\tau_c}+1\right)^{-p}\right]~~. 
  \label{deltar2_KVMH_uni}
\end{equation}

	In addition, the time-dependent diffusion coefficient $D(\tau)$, which can be evaluated through the derivative 
of the MSD (see, e.g., Eq.~\ref{D_KV_final}), is given by
\begin{equation}
  D(\tau) = \frac{k_B T}{\kappa\tau_c}\left(\frac{2\tau}{p\tau_c}+1\right)^{-(1+p)}~~.
  \label{D_final}
\end{equation}
	It is worth emphasizing that, at later times, the above result indicates that the time-dependent diffusion 
coefficient behaves like a power-law, that is, $D(\tau) \propto \tau^{-(1+p)}$.
	Essentially, this is the main feature that distinguishes the KVMH model from the usual KV model.
	Indeed, the distinction between an exponential, Eq.~\ref{D_KV_final}, and a power-law, Eq.~\ref{D_final}, 
can be observed in microrheological experiments, so it should be used to indicate the presence of 
micro-heterogeneities as the exponent $p$ is the quantity that defines the distribution $\rho(\xi)$ (Eq.~\ref{rho_dist}).

	One should note that, for large values of $p$, the negative term in the MSD given by Eq.~\ref{deltar2_KVMH_uni} 
goes to an exponential function, {\it i.e.}, $e^{-z}=\lim_{p \rightarrow \infty}(1 + z/p)^{-p}$, so that it recovers
the MSD obtained for the usual KV model (Eq.~\ref{deltar2_KV_uni}).
	Also, large $p$ leads the distribution $\rho(\xi)$ (Eq.~\ref{rho_dist}) to be more localized around a prominent peak, 
which means that only a particular value $\xi^{*}$ will contribute to the average of $\xi$,  fading out the effect of 
the micro-heterogeneities.

	In addition, it is worth noting that the MSD given by Eq.~\ref{deltar2_KVMH_uni} can be thought of as a particular case 
of a general result obtained in Ref.~\citep{rizzi2020jrheol}, which reads
\begin{equation}
  \langle \Delta x^2(\tau) \rangle = \frac{k_B T}{\kappa} \left\{1-\left[\left(\frac{\tau}{\tau^*}\right)^{n^*}+1\right]^{-\alpha}\right\},
  \label{deltar2_final_leandro}
\end{equation}
where $\tau^*$ is a characteristic time, $\alpha$ is a parameter related to a gamma distribution just like 
Eq.~\ref{rho_dist}, and $n^*$ is an exponent related to the nature of the diffusive behaviour that can be eventually 
observed at short times.
	As discussed in Ref.~\citep{rizzi2020jrheol}, where a generalized overdamped Langevin approach is used, the cases 
where $0.5<n^*<1$ denote a non-Markovian subdiffusive behaviour, while $n^*=1$ correspond to the case where the normal 
diffusive behaviour of the probe particles is Markovian, as for the KV and KVMH models (see Eqs.~\ref{short-times-final} 
and~\ref{deltar2_KVMH_uni}).

  Finally, it is important to note that, as also pointed out in Ref.~\citep{rizzi2020jrheol}, the use of a gamma 
distribution $\rho(\xi)$ (Eq.~\ref{rho_dist}) has been established because of the link between the effective elastic 
constants and the size of clusters of particles with a given size~\citep{KRALL199719,PhysRevLett.96.185502,zaccone2014jrheol}.
	Besides the relationship to the cluster sizes, gamma distributions are not only derived from kinetic 
equations~\citep{zaccone2014jrheol} which can be related to the relaxation times distributions $H(\lambda)$ observed 
experimentally from mechanical spectroscopy~\citep{winter2000chapter5}, but are also observed in the disordered networks 
generated by numerical simulations~\citep{hexner2018pre,hexner2018softmatter}.

%========================================================================
\subsection{Related distributions: elastic constants $\varepsilon$ and characteristic times $\lambda$}
\label{related_dist}
%========================================================================

  As it will become clear in the Sec.~\ref{numerical_sim}, the distribution $\rho(\xi)$ given by Eq.~\ref{rho_dist} can be
interpreted either in terms of local elastic constants $\varepsilon$ or, alternatively, in terms of local characteristic 
times $\lambda$, both which will be given by generalized gamma distributions~\citep{Crooks2019}.
	In particular, if we consider that $\varepsilon = (\kappa/p)\xi$, the distribution of local spring constants 
$\rho(\varepsilon)$ will be given by
$\rho(\varepsilon) =  \rho(\xi)|_{\xi=p\varepsilon/\kappa}\left(\partial \xi/\partial \varepsilon \right)$, that is,
\begin{equation}
  \rho(\varepsilon) =  \left(\frac{p}{\kappa}\right)^p\frac{\varepsilon^{p-1}}{\Upgamma(p)}\exp\left(-\frac{p\varepsilon}{\kappa}\right)~~,
  \label{elastic_constant_distribution}
\end{equation}
so that the average value of the effective spring constants will be given by
\begin{equation}
  \bar{\varepsilon} = \left( \frac{\kappa }{p} \right) \bar{\xi} = \kappa~~.
\end{equation}
  Furthermore, by considering that the local characteristic time is given by $\lambda=p\tau_c/\xi$, one can obtain the
distribution $H(\lambda)$ from the distribution $\rho(\xi)$ as
$H(\lambda) =  \rho(\xi)|_{\xi=p\tau_c/\lambda}\left( \partial \xi / \partial \lambda \right)$, which yields
\begin{equation}
  H(\lambda) = \frac{1}{\Upgamma(p)}\frac{1}{p\tau_c}\left(\frac{p\tau_c}{\lambda}\right)^{1+p}\text{exp}\left(-\frac{p\tau_c}{\lambda}\right)~~.
  \label{charac_times_dist}
\end{equation}
  Accordingly, in this last case, the average value of the characteristic times $\lambda$  is given by
\begin{equation}
  \overline{\lambda} = \int \lambda H(\lambda)d\lambda = \frac{\tau_c}{p^{-1}-1}~~.
\end{equation}
  We note that, although $\overline{\lambda}$ is different from the mean characteristic time $\tau_c$, the average
value of the reciprocal of the characteristic times, $\lambda^{-1}$, will be given in terms of $\tau_c$ as
\begin{equation}
\overline{\lambda^{-1}} = \int \frac{1}{\lambda} H(\lambda)d\lambda = \frac{1}{\tau_c}~~.
\end{equation}

%========================================================================
\subsection{Time-dependent van Hove distributions}
%========================================================================

 	To characterize the movement of probe particles, one can also compute the van Hove distributions, which
can be easily determined from experiments and gives the probability of a given displacement $x$ after a time 
interval $\tau$.
  Analytically, one can evaluate the position distributions $f(x,\tau)$ as
\begin{equation}
  f(x,\tau) = \int  f_\xi(x,\tau) \, \rho(\xi) \,d\xi ~~,
  \label{position_dist}
\end{equation}
where~\citep{azevedo2020jphysconfser}
\begin{equation}
  f_\xi(x,\tau)=\left(\frac{1}{2\pi \langle \Delta x^2(\tau) \rangle_{\xi}}\right)^{1/2}
    \exp\left(-\frac{x^2}{2\langle \Delta x^2(\tau) \rangle_{\xi}}\right)~~,
  \label{local_position_dist}
\end{equation}
corresponds to the position distribution observed from the random movement of a probe in a particular mesoscopic 
region of the sample.
	The normal distribution is typically expected for $f_\xi(x,\tau)$ because we are assuming that, locally, 
the random moves of a probe particle can be still described by a Gaussian process.

	Unfortunately, it is difficult to evaluate the full analytic expression for $f(x,\tau)$, i.e., for any arbitrary 
time interval $\tau$, from Eq.~\ref{position_dist} considering Eq.~\ref{local_position_dist} with the MSD given 
by Eq.~\ref{deltar2_KVMH_uni}.
	However, if we consider the approximation for the MSD at short times ($\tau \ll \tau_c$), that is, 
$\langle \Delta x^2(\tau) \rangle_{\xi} \approx  (2 k_BT\xi/\zeta p) \tau$, with $\xi$ being a random variable 
that follows the distribution $\rho(\xi)$ given by Eq.~\ref{rho_dist}, Eq.~\ref{position_dist} 
leads to~\citep{gradshteyn}
\begin{equation}
  f(x,\tau) = \frac{2C_\tau (C_\tau x)^{p-1/2}}{\sqrt{\pi}\Gamma(p)}K_{p-1/2}(2C_\tau |x|)~~,
  \label{short_time_position_distribution}
\end{equation}
where $C_\tau=(\zeta p/4k_BT\tau)^{1/2}$ and $K_{\phi}(x)$ is the modified Bessel function of the second kind
of order $\phi$.
	On the other hand, at later times, at the steady state ($\tau \gg \tau_c$), the MSD will be independent of 
$\xi$ and $\tau$ (see Eq.~\ref{later-times}), so that the van Hove distribution evaluated through 
Eqs.~\ref{position_dist} and~\ref{local_position_dist} should be given by a normal distribution as well, that is, 
\begin{equation}
  f(x) = \sqrt{\frac{\kappa}{2\pi k_BT}}\exp\left(-\frac{\kappa x^2}{2k_BT}\right) ~~.
  \label{final_postion_dist}
\end{equation}

\noindent

%========================================================================
%========================================================================
\section{Numerical simulations}
\label{numerical_sim}
%========================================================================
%========================================================================

	%-%-%-%-%-%-%-%-%-%%-%-%-%-%-%-%-%-%-%
	%-%-%-%-%-%-%-%-%-%%-%-%-%-%-%-%-%-%-%
	%-%-%-%-%-%-%-%-%-%%-%-%-%-%-%-%-%-%-%
	\begin{figure*}[!t]
		\centering
		\includegraphics[width=0.99\textwidth]{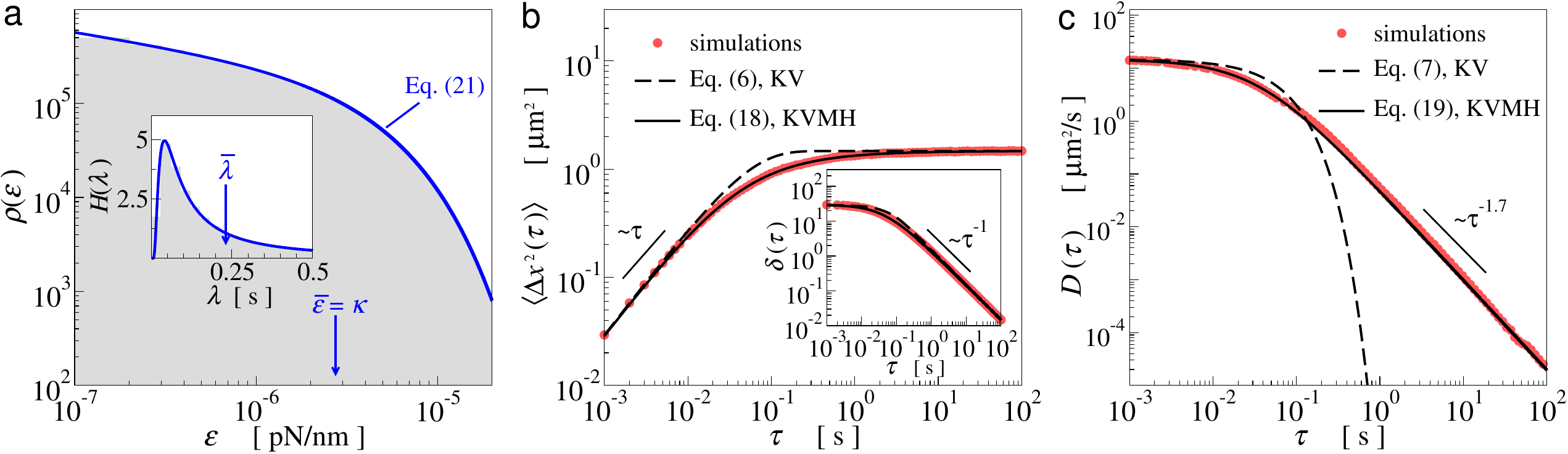}
		\caption{Numerical and theoretical results obtained for the KV and KVMH models. 
	    (a)~Histogram grey) and analytical distribution (continuous blue line, Eq.~\ref{elastic_constant_distribution}) 
	    of local elastic constants $\rho(\varepsilon)$ with $p=0.7$ and $\kappa=2.8 \times 10^{-6}\,$pN/nm.
    	Inset:~Histogram and analytical distribution (Eq.~\ref{charac_times_dist}) of characteristic times $H(\lambda)$.
    	(b)~Mean-squared displacement $\langle \Delta x^2 (\tau) \rangle$. Inset: Ratio 
    	$\delta (\tau)= \langle \Delta x^2 (\tau) \rangle/\tau$. 
	    (c)~Time-dependent diffusion coefficient $D(\tau)$.	Filled (red) circles correspond to results obtained from 
	    numerical simulations with $T=298$ K, $\kappa=2.8 \times 10^{-6}\,$pN/nm, $\zeta=0.28 \times 10^{-6}\,$pN.s/nm, 
	    $N_t=10^5$ trajectories and $\Delta \tau=10^{-4}\,$s, while the (black) lines denote results obtained from the KV 
	    model (dashed lines), Eqs.~\ref{deltar2_KV_uni} and~\ref{D_KV_final}, and the KVMH model (continuous lines),
      Eqs.~\ref{deltar2_KVMH_uni} and \ref{D_final} with $p=0.7$, where in the expressions for both models we assume 
      the same values for $T$, $\kappa$, and $\zeta$ as in the simulations.
}
		\label{fig_simulations}
	\end{figure*}
	%-%-%-%-%-%-%-%-%-%%-%-%-%-%-%-%-%-%-%
	%-%-%-%-%-%-%-%-%-%%-%-%-%-%-%-%-%-%-%
	%-%-%-%-%-%-%-%-%-%%-%-%-%-%-%-%-%-%-%

	As discussed in Ref.~\citep{azevedo2020jphysconfser}, the trajectories of the probe particles can be obtained 
through overdamped Brownian dynamics simulations by considering an Euler integration scheme which leads to the 
following discretized equation
\begin{equation}
  x_{i+1} = x_{i} - \frac{\kappa}{\zeta} x_i \, \Delta \tau + \sqrt{2 \frac{k_B T}{\zeta} \Delta \tau } ~ \text{N}(0,1)~~,
  \label{discretizedX}
\end{equation}
with $\Delta t=t_{i+1}-t_i$, $x_i=x(t_i)$, $x_0=0$, and $\text{N}(0,1)$ being a Gaussian variable with zero mean 
and variance equal to one.
	In order to consider the micro-heterogeneities present in the system, we assume a position-dependent drag 
coefficient $\nu_n = \nu(\vec{R}_n)$, so that Eq.~\ref{discretizedX} becomes
\begin{equation}
  x_{i+1} = x_{i} - \frac{\kappa}{\nu_n} x_i  \, \Delta \tau + \sqrt{2 \frac{k_B T}{\nu_n} \Delta \tau } ~ \text{N}(0,1)
  \label{discretized_Xn}
\end{equation}
Hence, the $n$-th particle will display a single trajectory that probes the $n$-th mesoscopic region within the 
sample (see, e.g., Fig.~\ref{referenceframe}).
	Here, just as in the relationship established by Eq.~\ref{nu-xi=p_xi_zeta}, we consider that
\begin{equation}
  \nu_n = p \frac{\zeta}{\xi_n} ~~,
  \label{nu_xi}
\end{equation}
with $\xi_n$ being the $n$-th random variable sampled according to a gamma distribution given by Eq.~\ref{rho_dist}.
	Hence, with this choice of $\nu_n$, the discretized Langevin equation, Eq.~\ref{discretized_Xn}, becomes
\begin{equation}
  x_{i+1} = x_{i} - \left( \frac{1}{p} \frac{\kappa}{\zeta} \xi_n \right)  x_i  \, \Delta \tau +
    \sqrt{2 \frac{k_B T}{\kappa} \left( \frac{1}{p} \frac{\kappa}{\zeta} \xi_n \right) \Delta \tau }  ~ \text{N}(0,1)~~,
  \label{discretized_xi}
\end{equation}
which can be used to simulate the $n$-th distinct mesoscopic region within the sample that displays slightly 
different viscoelastic properties.
	Interestingly, one can also interpret Eq.~\ref{discretized_xi} as an expression for a medium that displays different 
local spring constants which are given by
\begin{equation}
  \varepsilon_n = \left( \frac{\kappa}{p} \right) \xi_n ~~,
\end{equation}
just as defined in Section~\ref{related_dist}.
	Besides, it is worth noting that, based on Eq.~\ref{discretized_xi}, one has that the time-correlation function 
of the forces acting at the $n$-th probe particle can be written as~\citep{gillespie1993amjphys}
\begin{equation}
  \langle \tilde{f}_a(\tau) \tilde{f}_a(\tau^{\prime}) \rangle_{n}^{\,} = 
    2 \frac{k_BT}{\kappa} \frac{\varepsilon_{n}}{\zeta} \delta(\tau - \tau^{\prime})~~,
\end{equation}
which means that the fluctuation-dissipation theorem is locally valid, where {\it local} here means that the probe's 
trajectory is spatially localized within a mesoscopic region which is characterized by a single value of 
$\varepsilon_{n}$.

	Figure~\ref{fig_simulations} presents results obtained from numerical simulations that not only validate the 
algorithm we have just described but also demonstrate the theoretical results obtained from the KVMH model, 
illustrating its main differences from the usual KV model.
	In Fig.~\ref{fig_simulations}(a) we include the distribution of local spring constants $\rho(\varepsilon)$, 
Eq.~\ref{elastic_constant_distribution}, and the corresponding distribution of characteristic times $H(\lambda)$, 
Eq.~\ref{charac_times_dist}, that were used to perform the simulations.
	Figure~\ref{fig_simulations}(b) shows the MSDs obtained for the KV and KVMH models, given by 
Eqs.~\ref{deltar2_KV_uni} and~\ref{deltar2_KVMH_uni}, respectively.
	Accordingly, the two models display the same asymptotic behaviours for short and long times, where we observe 
a linear behaviour, $\langle \Delta x^2(\tau) \rangle \approx 2 (k_BT / \zeta) \tau$, and a plateau, 
$\langle \Delta x^2(\tau) \rangle \approx k_BT / \kappa$, respectively. 
	Also, as the numerical results presented in the inset panel in (b) indicate, the ratio 
$\delta(\tau)=\langle \Delta x^2(\tau) \rangle/\tau$ follows the expected behaviour according to the analytical 
results denoted by the black lines.
	For instance, for the KVMH, $\delta(\tau)$ is given by Eq.~\ref{short-times-final} divided by $\tau$, so it is
constant at short times, and, at later times, it displays an algebraic decay proportional to $\tau^{-1}$, with 
$\delta(\tau)$ given by a constant, i.e., Eq.~\ref{later-times-final}, divided by $\tau$.

	Even though the difference between the KV and KVMH models can be noticed from the results obtained for the MSD 
at intermediate times, the difference is better observed in the behaviour of the time-dependent diffusion 
coefficient $D(\tau)$.
	As shown in Fig.~\ref{fig_simulations}(c), the KV model displays an exponential decay, Eq.~\ref{D_KV_final}, 
while the KVMH model exhibits a power-law behaviour, Eq.~\ref{D_final}, with $D(\tau) \propto \tau^{-1.7}$ for 
$p=0.7$.
	We note that, in practice, the derivative used to compute the diffusion coefficient was numerically determined 
from the MSD through the expression $D(\tau)=\langle \Delta x^2(\tau) \rangle p(\tau)/2\tau$, where 
$p(\tau)=\text{dln}\langle \Delta x^2(\tau) \rangle/\text{dln}\tau$ is the effective exponent of the MSD, obtained 
here through a numerical linear regression.
	It is worth mentioning that the observed power-law behaviour of $D(\tau)$ at later times is in agreement with 
the experimentally observed behaviour of certain hydrogels~\citep{teixeira2007jphyschemB}.

%-%-%-%-%-%-%-%-%-%%-%-%-%-%-%-%-%-%-%
%-%-%-%-%-%-%-%-%-%%-%-%-%-%-%-%-%-%-%
%-%-%-%-%-%-%-%-%-%%-%-%-%-%-%-%-%-%-%
\begin{figure}[!t]
	\centering
	\includegraphics[width=0.42\textwidth]{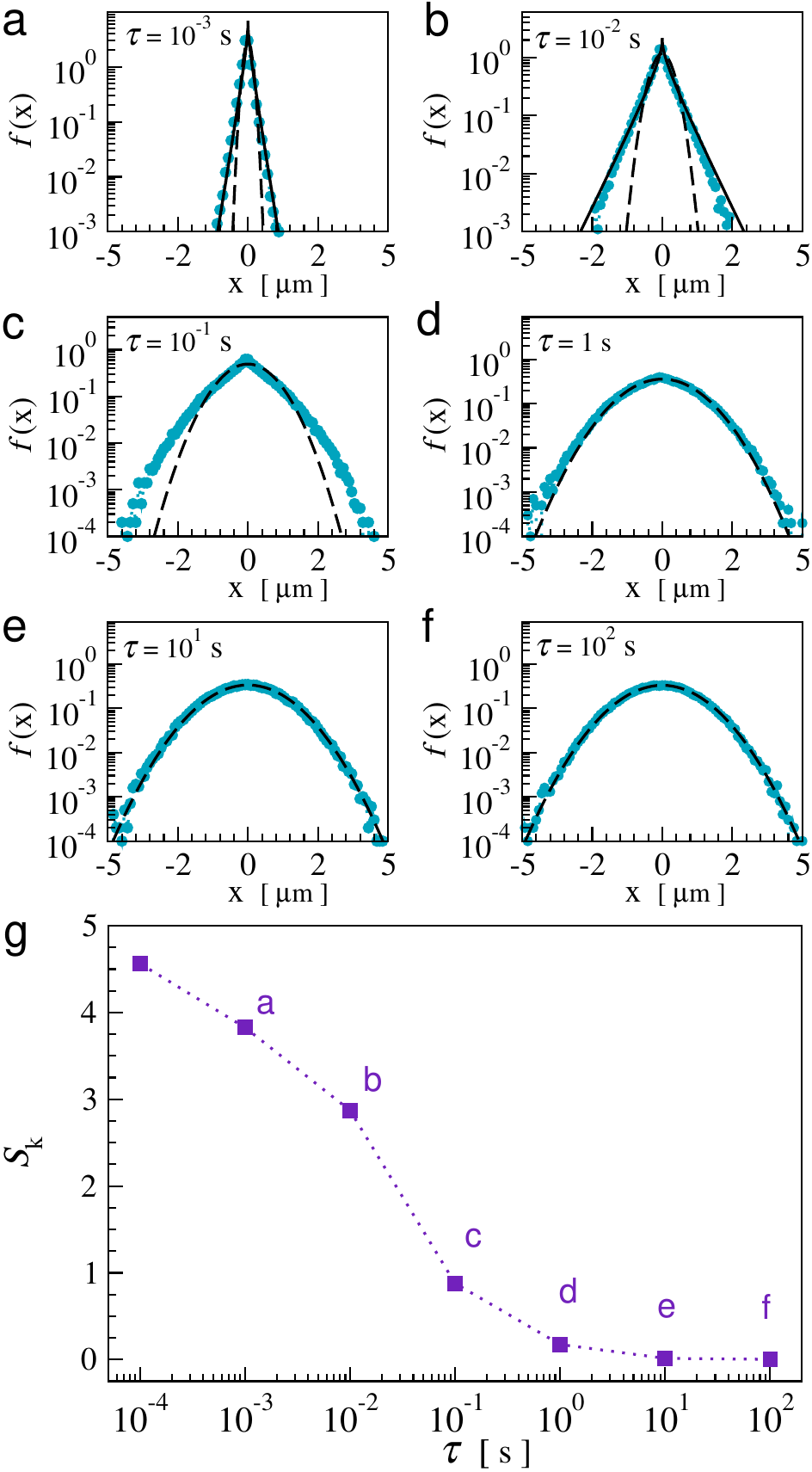}
	\caption{(a)-(f) Position distributions $f(x,\tau)$ at different times $\tau$ determined for the KVMH model.
      Filled symbols correspond to histograms evaluated from $N_t=10^5$ trajectories obtained from numerical simulations
      in one dimension ($d_e=1$) as described in Sec.~\ref{numerical_sim}, with $T=298\,$K, $\kappa=2.8 \times 10^{-6}\,$pN/nm, 
      $\zeta=0.28 \times 10^{-6}\,$pN.s/nm, $p=0.7$, and $\Delta \tau=10^{-4}\,$s. Continuous black lines denote the expected 
      analytical distribution for short times given by Eq.~\ref{short_time_position_distribution}, while dashed black lines 
      correspond to Gaussian distributions, Eq.~\ref{final_postion_dist}.
	    (g) Excess kurtosis $S_{\text{k}}(\tau)$ given by Eq.~\ref{excess-kurtosis}.
	    }
	\label{fig_distributions}
\end{figure}
%-%-%-%-%-%-%-%-%-%%-%-%-%-%-%-%-%-%-%
%-%-%-%-%-%-%-%-%-%%-%-%-%-%-%-%-%-%-%
%-%-%-%-%-%-%-%-%-%%-%-%-%-%-%-%-%-%-%

	In Fig.~\ref{fig_distributions}(a)-(f) we show the van Hove distributions $f(x,\tau)$ obtained numerically from the 
simulations at different times $\tau$, together with the expected analytical distribution given by 
Eq.~\ref{short_time_position_distribution} at short times (continuous lines) and the Gaussian distributions predicted 
by the KV model defined in Sec.~\ref{KV_section} given by Eq.~\ref{final_postion_dist} (dashed lines).
	In order to generate the Bessel function $K_{\phi}(x)$ we used the built-in function {\ttfamily scipy.special.kv} of 
the Scipy library~\citep{scipy}.
	For small orders $\phi$ and for $x>0$, the algorithm uses the following integral representation~\citep{gradshteyn}
\begin{equation}
	K_{\phi}(x)=\text{sec}\left(\frac{ \phi  \pi}{2}\right)\int_{0}^{\infty}\text{cos}(x\, \text{sinh}(t'))\, \text{cosh}( \phi \, t')\,dt',
\end{equation}
for $-1<\text{Re}\{\phi\}<1$.

	Figure~\ref{fig_distributions} indicates that the distributions obtained at short times are similar to 
Eq.~\ref{short_time_position_distribution} and approach Gaussian distributions at later times.
	In order to assess the deviation of the obtained distributions from Gaussian distributions, we obtain 
the excess kurtosis which is defined as~\citep{donald2008eurphysJE}
\begin{equation}
  S_{\text{k}}(\tau)=\sum_{j=1}^{N_t} \frac{[x_j(\tau)-\bar{x}(\tau)]^4}{(N_t-1) [\sigma_x(\tau)]^4}-3,
  \label{excess-kurtosis}
\end{equation}
where $N_t$ is the number of trajectories, $\bar{x}$ is the mean and $\sigma_x$ is the standard deviation of the distribution.
	Figure~\ref{fig_distributions}(g) shows that $S_{\text{k}}(\tau)$ obtained for the distributions at different times is higher 
for short times, but it approaches zero at later times, as the excess kurtosis is expected to be zero for Gaussian distributions.

	A non-zero excess of kurtosis at short times $\tau$ is precisely what is observed during sol-gel transition gelation examined 
through microrheology experiments~\citep{donald2008eurphysJE,oppong2008,rich2011}. 
	In those experiments, a zero excess of kurtosis is observed for short gelation times at any time $\tau$ (characterizing the sol phase of the samples) followed by an increase in excess kurtosis as the sample jellifies and becomes more heterogeneous 
(characterizing their gel phase).
	In these cases, the increase in the excess kurtosis can be clearly associated with the presence of heterogeneous structures in the semisolid materials.

%========================================================================
%========================================================================
\section{Application of the KVMH model to experimental data}
\label{Experimental_section}
%========================================================================
%========================================================================

	In this section, we present a comparison between the expressions obtained from the KVMH model and some experimental data to further validate our approach.

\subsection{Results for polyacrylamide gels from Ref.~\citep{Dasgupta2005}}

\begin{figure}[!t]
	\centering
	\includegraphics[width=0.48\textwidth]{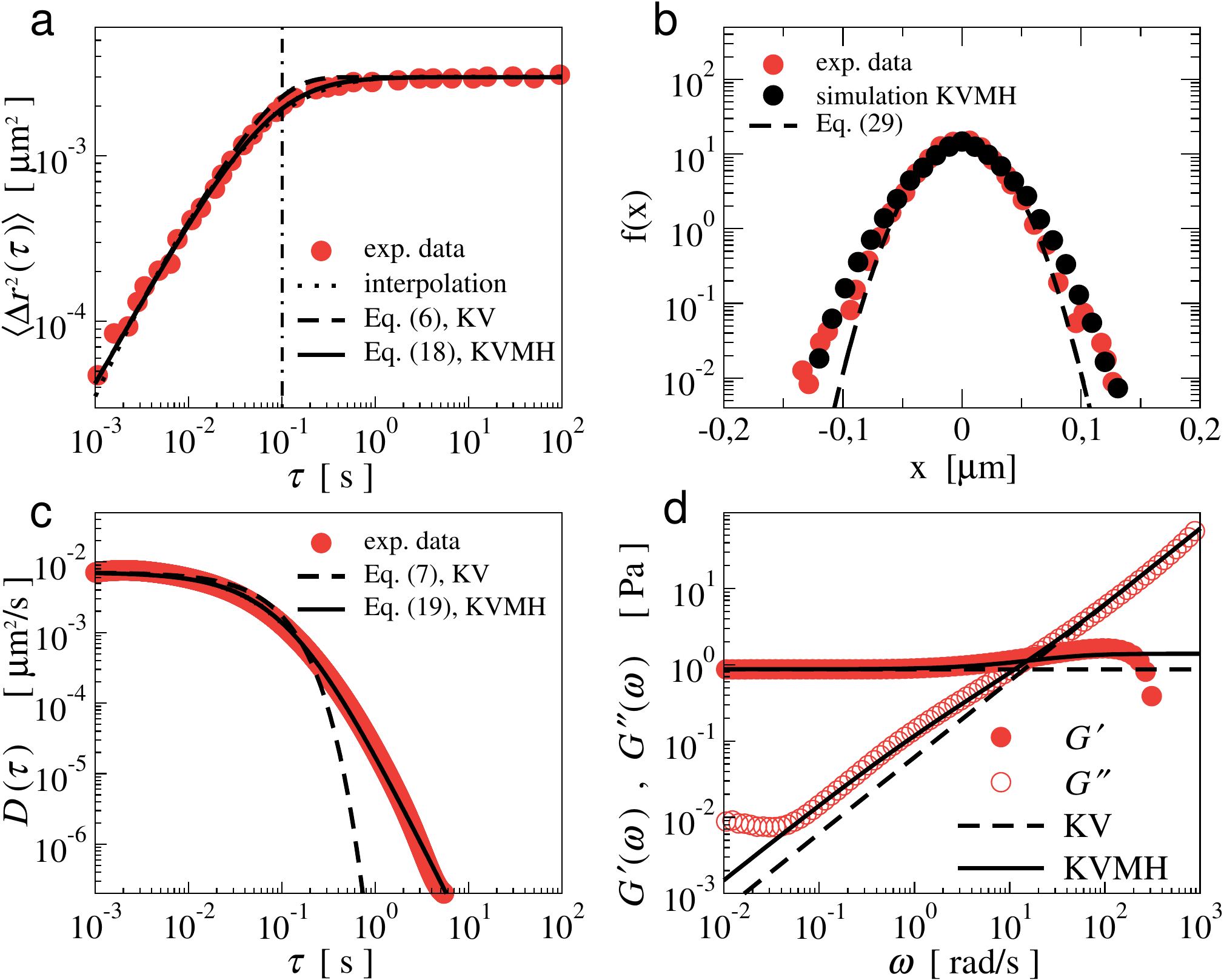}
	\caption{Comparison between the experimental data for polyacrylamide  gels (filled red circles) extracted from 
	        Ref.~\citep{Dasgupta2005} and the results obtained from the usual KV model with $T=298\,$K, 
	        $\kappa=2.75\,$pN/$\mu$m, $\tau_c=0.14\,$s, and the KVMH with the same parameters but a distribution 
	        characterized by $p=1.67$. 
          (a)~MSD $\langle \Delta r^{2}(\tau) \rangle$.	The vertical dash-dotted line indicates $\tau=0.1\,$s.
          (b)~ Position distributions $f(x,\tau)$ at time $\tau=0.1\,$s. Filled red and black circles denote the 
          distributions obtained from experiments and numerical simulations, respectively, while the dashed
          line corresponds to a Gaussian distribution, Eq.~\ref{final_postion_dist}.
          (c)~Time-dependent diffusion coefficient $D(\tau)$ (here the experimental estimates were obtained from a
          smoothed interpolation~\citep{cubic_spline} using Eq.~\ref{interpolation} with $\varphi=0.003$).
          (d)~Storage modulus $G^{\prime}(\omega)$ (filled circles) and loss modulus $G^{\prime \prime}(\omega)$ 
          (open circles), obtained from $\langle \Delta r^{2}(\tau) \rangle$ via Eqs.~\ref{compliance} 
          and~\ref{complemodulus}, with $\hat{J}(\omega)$ obtained through Eq.~\ref{transfj} using the numerical 
          method described in Ref.~\citep{Evans2009}.}
	    \label{fig_exp_poli_weitz}
\end{figure}

	We first include in Fig.~\ref{fig_exp_poli_weitz} the comparison between the KV and KVMH models and the experimental data 
extracted from Ref.~\citep{Dasgupta2005}, where microrheology experiments were carried out  for chemically cross-linked 
polymer polyacrylamide gels.
	The experiments were done with the particle tracking technique~\citep{Rizzi_Tassieri_2018} using positively charged aldehyde
amidine polystyrene spheres with radius $a=0.05\,\mu$m at room temperature ($T=298\,$K).

	As it can be seen in Fig.~\ref{fig_exp_poli_weitz}(a), the experimental data obtained for the MSD presents an approximately 
linear behaviour at short times and a plateau at long times, just as the expressions derived for the KV and the KVMH models, 
with the latter showing a slightly better agreement with the experimental data.
	In Fig.~\ref{fig_exp_poli_weitz}(b) we include the experimentally determined~\citep{Dasgupta2005} van Hove distribution 
$f(x,\tau)$ at time $\tau=0.1\,$s, together with the distribution obtained from numerical simulations using the parameters 
obtained from the fit of the MSD determined for the KVMH model displayed in \ref{fig_exp_poli_weitz}(a).
  Accordingly, the good agreement between the numerical and experimental distributions $f(x,\tau)$ in 
Fig.~\ref{fig_exp_poli_weitz}(b), including their matching non-Gaussianity, serves as an indirect validation of the gamma 
distribution, Eq.~\ref{rho_dist}, used to define the KVMH.

	Figure~\ref{fig_exp_poli_weitz}(c) includes results for the time-dependent diffusion coefficient $D(\tau)$.
	It is worth mentioning that, due to the noisy nature of the MSD data, the diffusion coefficient $D(\tau)$ was obtained from 
a smoothed interpolation of the experimental data (see Ref.~\citep{cubic_spline} for a description of the algorithm).
	The idea is that, given a sequence of $N_k$ measurements $\{v_1,v_2,\cdots,v_{N_k}\}$, one minimizes
\begin{equation}
	\varphi \sum_{j=1}^{N_k}[v_j(\tau)-h_j(\tau)]^2+\sum_{j=1}^{N_k-2}[h_{j+2}(\tau)-2h_{j+1}(\tau)+h_j(\tau)],
  \label{interpolation}
\end{equation}
where $h_j(\tau)$ are the values that interpolates the data and $\varphi$ is a positive real smoothing parameter.
	The larger $\varphi$ the more the solution converges to the measured sequence.
	The smaller $\varphi$ the smoother the interpolation.
	Here the values $\varphi$ were chosen in a way that $D(\tau)$ does not display any oscillatory behaviour.
	The comparison to experimental results in Fig.~\ref{fig_exp_poli_weitz}(c) evidences the difference between the models' 
predictions, from where one can see that the exponential decay of $D(\tau)$ predicted by the KV model does not fit 
satisfactorily the experimental data very well.
	Conversely, the value of $p = 1.67$ determined from $D(\tau)$ given by the KVMH suggests that the presence of 
micro-heterogeneities in this gel are relevant, reinforcing the results presented in Fig.~\ref{fig_exp_poli_weitz}(b).

%-%-%-%-%-%-%-%-%-%%-%-%-%-%-%-%-%-%-%
%-%-%-%-%-%-%-%-%-%%-%-%-%-%-%-%-%-%-%
%-%-%-%-%-%-%-%-%-%%-%-%-%-%-%-%-%-%-%
%-%-%-%-%-%-%-%-%-%%-%-%-%-%-%-%-%-%-%
%-%-%-%-%-%-%-%-%-%%-%-%-%-%-%-%-%-%-%
%-%-%-%-%-%-%-%-%-%%-%-%-%-%-%-%-%-%-%

	Finally, we include in Fig.~\ref{fig_exp_poli_weitz}(d) a comparison to show how the models perform in describing the 
behaviour of the shear moduli $G^{\prime}(\omega)$ and $G^{\prime \prime}(\omega)$.
	As mentioned in Sec.~\ref{KV_section}, Eqs.~\ref{compliance} and~\ref{complemodulus} allow us to obtain the complex 
shear modulus $G^{*}(\omega)$ of the material from the MSD.
	As we were not able to derive exact analytic expressions for the complex modulus $G^{*}(\omega)$ of the KVMH, we 
numerically compute the Laplace-Fourier transform $\hat{J}(\omega)$ through~\citep{Evans2009}
\begin{equation}
\begin{split}
  -\omega^2\hat{J}(\omega) = i\omega J(0)+(1-e^{-i\omega \tau_{1}})\frac{(J_{1}-J(0))}{\tau_{1}} \\ 
  + \dot{J}_{\infty}e^{-i\omega \tau_{N}}
  +\sum_{k=2}^{N}\left( \frac{J_{k}-J_{k-1}}{\tau_{k}-\tau_{k-1}}\right)(e^{-i\omega \tau_{k-1}}-e^{-i\omega \tau_{k}}),
\label{transfj}
\end{split}
\end{equation}
where one can assume that $J(0) = \text{lim}_{\tau \rightarrow 0} J(\tau) \,=0$ and $\dot{J}_{\infty} = \text{lim}_{\tau \rightarrow \infty}
dJ(\tau)/d\tau \,=0$. 
	Figure~\ref{fig_exp_poli_weitz}(d) indicates that the shear moduli obtained from the KV model deviate from the experimental data, especially at low frequencies for the loss modulus $G^{\prime \prime}(\omega)$ and at high frequencies 
for the storage modulus $G^{\prime}(\omega)$, while the KVMH yields a better correspondence.

\subsection{Dynamic light scattering (DLS) experiments}

	Next, we present results obtained from dynamic light scattering (DLS) experiments performed by our group. 
	Hence, in this Section, we include a brief description of the methods and experimental setup.
	The DLS measurements were made using a laser with a wavelength equal to $\lambda^{\prime}=632.8\,$nm, and the
normalized intensity correlation function $g^{(2)}(\tau)=\langle I(0) I( \tau) \rangle/\langle I^{2}\rangle$ of 
the scattering intensity $I(\tau)$ was obtained through a multi-angle detection system by Brookhaven Co.
with a TurboCo correlator.
	To ensure the proper ensemble average, a large pinhole($400\,\mu$m of diameter) was placed before the photon detector so the final average was a result of different positions across the samples.
	The MSD of the probe particles was estimated experimentally as~\citep{teixeira2007jphyschemB}
\begin{equation}
  \left\langle \Delta r^{2} \left( \tau \right)  \right\rangle =  
    -\frac{6}{q^{2}} \text{ln}\left( \frac{g^{(2)}(\tau) - 1 - \sigma + \beta}{\beta} \right)^{1/2},
  \label{MSD-from-g2}
\end{equation}
where $q = (4 \mu \pi/ \lambda^{\prime} )\text{sin}\left( \theta/2 \right)$ is the modulus of the scattering 
vector, which is obtained from the wavelength $\lambda'$, the refraction index of water $\mu = 1.331$, and the 
scattering angle $\theta$.
	The parameters $\beta$ and $\sigma$ in Eq.~\ref{MSD-from-g2} correspond to the extrapolation of the autocorrelation 
function for $\tau \rightarrow 0$ for homodyne scattering (in this case, standard polystyrene particles in aqueous 
solution) and heterodyne scattering (polystyrene particles in the gels), respectively~\citep{bernepecora}.

\subsection{Gelation process in laponite gels}

	First, to illustrate the application of the KVMH model in the description of a very heterogeneous gel, 
we include here measurements related to the gelation process of laponite gels.
	In particular, we considered 10\,mL of laponite hydrogel synthesized in 3$\%$ m/m concentration, together 
with 5\,$\mu$L of solution with polystyrene particles with radius $a=0.5\,\mu$m (Spherotech). 
	The sample was stirred and sonicated for some minutes to mix the laponite in the solution.
	The measurements were made at $T=298\,$K along 136 hours, with accumulation times of one hour to obtain $g^{(2)}(\tau)$ and for a scattering angle equal to 90$^{\circ}$.

\begin{figure}[!t]
	\centering
	\includegraphics[width=0.48\textwidth]{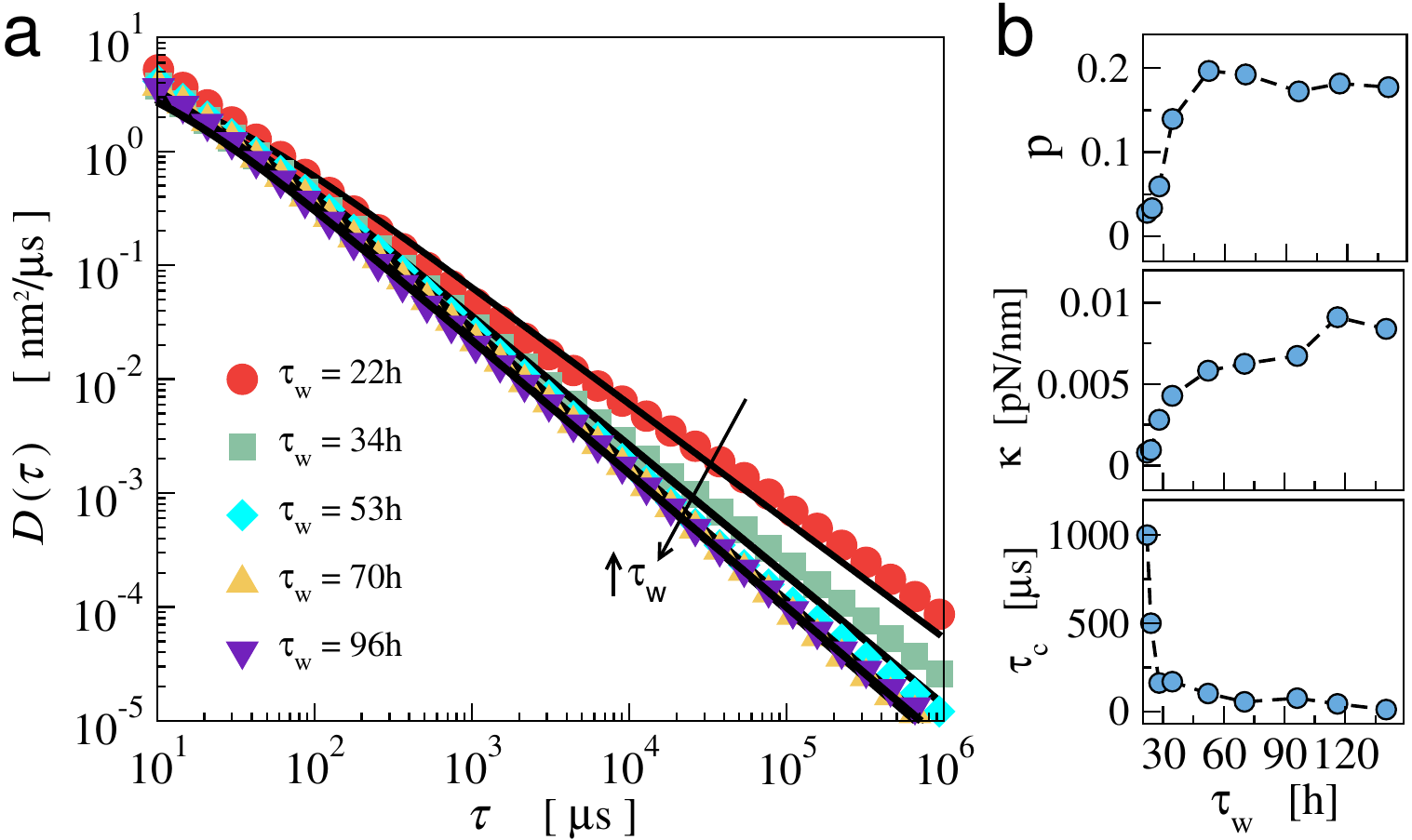}
	\caption{(a)~Time-dependent diffusion coefficient $D(\tau)$ of polystyrene particles ($a=0.5\mu$m) during 
          the gelation of a Laponite gel in room temperature ($T=298\,$K) at different gelation times $\tau_w$. 
          (b)~Parameters $p$, $\kappa$, and $\tau_c$ obtained through the of fit of $D(\tau)$ obtained for 
          the KVMH model (continuous black lines, Eq.~\ref{D_final}) to the experimental data shown in (a) as 
          a function of $\tau_w$.
}
	\label{fig_exp_laponite}
\end{figure}

	Figure~\ref{fig_exp_laponite}(a) exhibits the time-dependent diffusion coefficient $D(\tau)$ for different 
gelation times $\tau_w$.
	The results show the decrease of the diffusion coefficient as $\tau_w$ increases, indicating further
entrapment of the particles in the microstructures of the gel.
	Clearly, the ageing of the gel is marked by a slightly decrease in the material's heterogeneity, as it can be 
inferred from the increase in the value of $p$ observed in Fig.~\ref{fig_exp_laponite}(b).
	Interestingly, previous research has established that this gel is composed of micron-sized aggregates
exhibiting fractal behaviour~\citep{Pignon1996,Pignon1997}. 
	Indeed, in this case, one can relate the exponent $p$ to the spectral dimension $d_s$ that characterizes 
the network topology~\citep{BenAvraham2002}, since the time-dependent diffusion coefficient is expected to 
behave as~\citep{teixeira2007jphyschemB} $D(\tau) \propto \tau^{-d_s/2}$ at long times. 
	In this regime, Eq.~\ref{D_final} gives $D(\tau) \propto \tau^{-(1+p)}$, so $p$ and $d_s$ are expected 
to be related as $d_s=2(1+p)$. 
	In particular, the saturation value of $0.2$ observed for of $p$  in Fig.~\ref{fig_exp_laponite}(b)
corresponds to $d_s=2.4$, which is a value that is close to that observed for different kinds of 
gels~\citep{teixeira2007jphyschemB}.

	Here it is worth noting that the fit of the expressions obtained from the KVMH model to the experimental
data also provides estimates for $\kappa$ and $\tau_c$.
	The increase in the material stiffness can be inferred from the increase in $\kappa$ in 
Fig.~\ref{fig_exp_laponite}(b), and	the observed decay in $\tau_c$ reflects the further entrapment of the
probe particles.

%-%-%-%-%-%-%-%-%-%%-%-%-%-%-%-%-%-%-%
%-%-%-%-%-%-%-%-%-%%-%-%-%-%-%-%-%-%-%
%-%-%-%-%-%-%-%-%-%%-%-%-%-%-%-%-%-%-%
%-%-%-%-%-%-%-%-%-%%-%-%-%-%-%-%-%-%-%
%-%-%-%-%-%-%-%-%-%%-%-%-%-%-%-%-%-%-%
%-%-%-%-%-%-%-%-%-%%-%-%-%-%-%-%-%-%-%

%-%-%-%-%-%-%-%-%-%%-%-%-%-%-%-%-%-%-%
%-%-%-%-%-%-%-%-%-%%-%-%-%-%-%-%-%-%-%
%-%-%-%-%-%-%-%-%-%%-%-%-%-%-%-%-%-%-%
\begin{figure}[!t]
	\centering
	\includegraphics[width=0.48\textwidth]{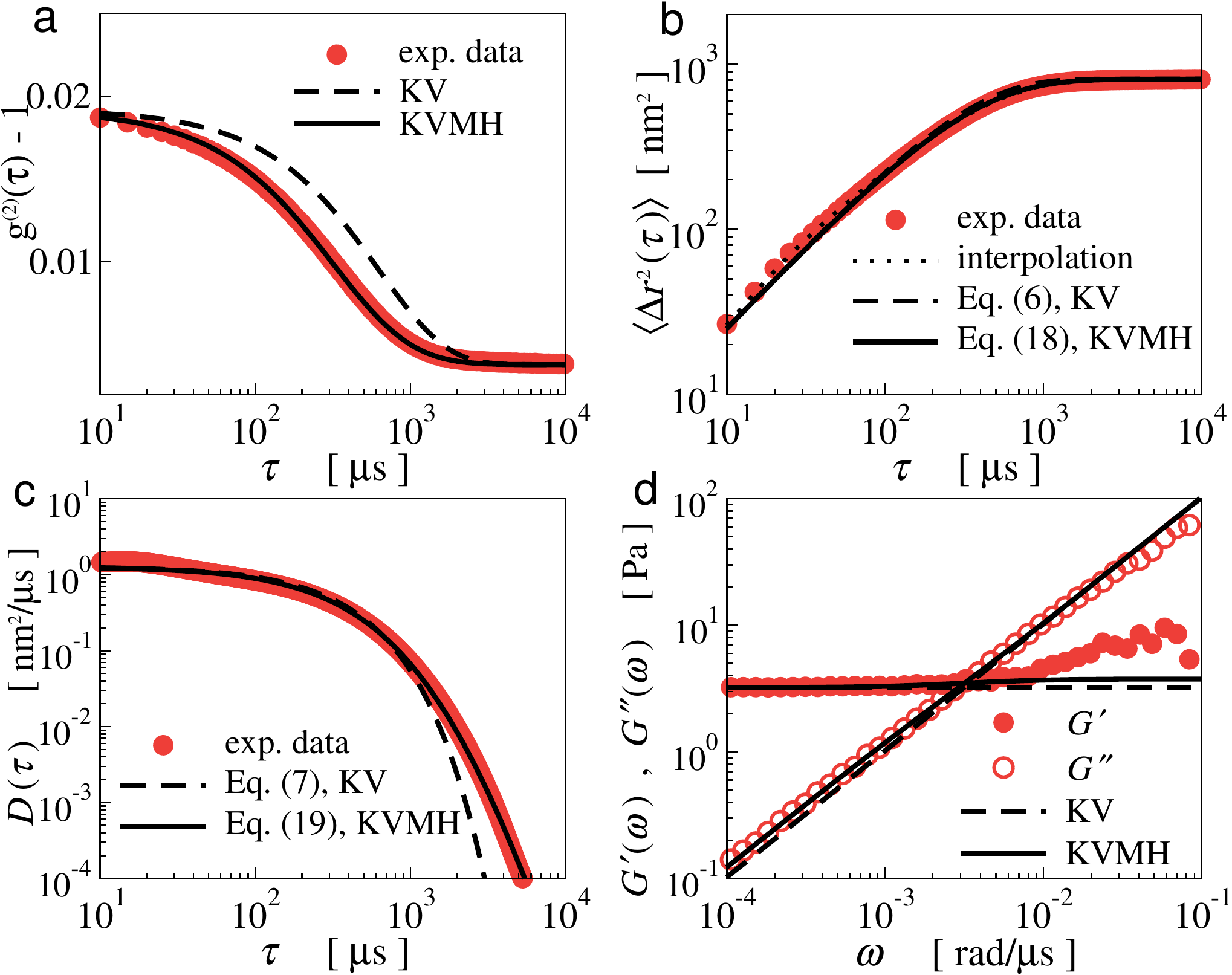}
	\caption{Comparison between the experimental data obtained for a polyacrylamide gel (filled red circles)
          and the theoretical expressions from the usual KV model with $T=298\,$K, $\kappa=1.52 \times 10^{-2}\,$pN/nm, 
          $\tau_c=636.47\,\mu$s, and the KVMH model with the same parameters and $p=5.98$. 
          (a)~Intensity-normalized correlation function $g^{(2)}(\tau)-1$. 
          (b)~MSD $\langle \Delta r^{2}(\tau) \rangle$ after 24h.
        	(c)~Time-dependent diffusion coefficient $D(\tau)$ (here the experimental estimates were obtained from a
          smoothed interpolation~\citep{cubic_spline} using Eq.~\ref{interpolation} with $\varphi=0.4$).
          (d)~Storage modulus $G^{\prime}(\omega)$ (filled circles) and loss modulus $G^{\prime \prime}(\omega)$ (open circles),
          obtained from $\langle \Delta r^{2}(\tau) \rangle$ via Eqs.~\ref{compliance} and~\ref{complemodulus}, with 
          $\hat{J}(\omega)$ obtained through Eq.~\ref{transfj} using the numerical method described in Ref.~\citep{Evans2009}.
}
  	\label{fig_exp_poli}
\end{figure}
%-%-%-%-%-%-%-%-%-%%-%-%-%-%-%-%-%-%-%
%-%-%-%-%-%-%-%-%-%%-%-%-%-%-%-%-%-%-%
%-%-%-%-%-%-%-%-%-%%-%-%-%-%-%-%-%-%-%

\subsection{Heterogeneity-related effects in polyacrylamide gels}

  Finally, to show a case where the semisolid material is not so heterogeneous, we present 
additional results that were obtained for a polyacrylamide gel.
	The gel was synthesized using acrylamide (A8887), bis-acrylamide (M7279), 
tetramethylethylenediamine (TEMED, T7024) and ammonium persulfate (APS, A3678), all purchased from SIGMA-ALDRICH.
	The synthesis was made at room temperature ($T\cong 298\,$K) using the ratio of acrylamide and 
bis-acrylamide equal to $30:1$, with total concentration equal $4.14\%\,$m/m.
	After this was added $5\,\mu$L of solution with polystyrene particles with radius $a=0.5\,\mu$m.
	To start the reaction, $5\,$mL of APS solution with $0.404\%\,$m/m and $10\,\mu$L of TEMED was added. 
	The sample was left to react at 0$^{\circ}$C overnight.
	The measurements were made in room temperature, i.e., $T=298\,$K, with 3 hours of duration, 6 accumulations 
(average over 6 measurements) at different sample positions and for a scattering angle equal to 40$^{\circ}$.
	The pinhole was set, as before, to $400\,\mu$m.
 
	Figure~\ref{fig_exp_poli} shows that although the KVMH model is still more adequate to describe the experimental 
results, the discrepancy between the fits of the two models to the experimental data is rather small.
	This occurred because of the high value of $p=5.98$.
	As $p\gg 1$, the distribution of local spring constants $\rho(\varepsilon)$, Eq.~\ref{elastic_constant_distribution}, 
becomes narrower with the more prominent peak around the mean $\bar{\varepsilon}=\kappa$.
	This is precisely the case of the semisolid Kelvin-Voigt described in Sec.~\ref{KV_section}, where the material is 
characterized by just one spring constant $\kappa$.
	Thus, this result indicates that the polyacrylamide gel produced in our laboratory is reasonably homogeneous.

%========================================================================
%========================================================================
\section{Concluding remarks}
\label{Concluding_section}
%========================================================================
%========================================================================

	In this work, we established a theoretical approach that generalizes the Kelvin-Voigt model 
to take into account the micro-heterogeneities observed in semisolid viscoelastic 
materials. 
	Our model naturally introduces the micro-heterogeneities in the context of microrheology, 
with the average over the various mesoscopic regions theoretically performed here being equivalent 
to the experimental measurements using the multiple particle tracking technique~\citep{Rizzi_Tassieri_2018}.

	By considering both experiments and simulations we were able to validate our theoretical results, 
including the expressions for the MSD (Eq.~\ref{deltar2_KVMH_uni}), for the time-dependent diffusion 
coefficient (Eq.~\ref{D_final}), and for the position distributions (Eq.~\ref{short_time_position_distribution} 
for short times and Eq.~\ref{final_postion_dist} for later times).
	Interestingly, our numerical and analytical findings indicate a direct relationship between the 
micro-heterogeneities of local spring constants, the non-Gaussianity of the van Hove distributions 
$f(x,\tau)$, and the power-law behaviour of $D(\tau)$ at long times.
	Hence, the obtained expressions for $\langle \Delta x^2(\tau)\rangle$ and $D(\tau)$ can be used by experimental
rheologists to yield estimates for the exponent $p$, so they provide a way to infer the relevance of the 
micro-heterogeneities to the response of KV-like semisolid materials.

	Finally, it is worth noting that although we were not able to derive exact analytical expressions for the 
shear moduli $G^{\prime}(\omega)$ and $G^{\prime \prime}(\omega)$ for the KVMH model, we showed that both moduli 
can be obtained numerically from the fit of our expressions to the experimental data through the method described 
in Ref.~\citep{Evans2009}.
	This model-fitting procedure allows one to extract reliable estimates for the shear moduli even when the MSD displays noisy behaviour at long times, as it usually occurs to experimental data obtained from microrheology techniques.

\vspace{0.5cm}

%\section*{Acknowledgements}
\noindent
{\bf Acknowledgements:}
T. N. Azevedo thanks the scholarship from CAPES, and L. G. Rizzi acknowledges the support from the brazilian agency CNPq (Grant N$^{\circ}$ 312999/2021-6).
The authors also thank the computational resources made available by GISC-UFV.

\vspace{0.5cm}

%\section*{Conflicts of interest}
\noindent
{\bf Conflicts of interest:}
There are no conflicts to declare.

\vspace{0.5cm}

%\section*{Data availability}
\noindent
{\bf Data availability:}
The data that support the findings of this study are available from the corresponding author upon reasonable request.

\vspace{0.5cm}

%\section*{Author Contributions}
\noindent
{\bf Author Contributions:}
{\bf T. N. Azevedo.} 
Investigation.
Data curation.
Formal analysis.
Software.
Writing - original draft.
Writing - review \& editing.
{\bf K. M. Oliveira.} 
Investigation.
Data curation.
Formal analysis.
Writing - original draft.
{\bf H. P. Maia.} 
Investigation.
Data curation.
Formal analysis.
{\bf A. V. N. C. Teixeira.} 
Investigation.
Data curation.
Formal analysis.
Methodology.
Supervision.
Validation.
Writing - review \& editing.
{\bf L. G. Rizzi.} 
Conceptualization.
Investigation.
Data curation.
Formal analysis.
Funding acquisition.
Methodology.
Software.
Visualization.
Project administration.
Supervision.
Validation.
Writing - original draft.
Writing - review \& editing.

\section*{References}

\vspace{-0.5cm}

%%%REFERENCES%%%
%\bibliography{biblio}


\end{document}